\documentclass[aps,prb,twocolumn,amsmath,amssymb,gallery,footinbib,superscriptaddress,floatfix]{revtex4-2}
\usepackage{graphicx}
\usepackage{dcolumn}
\usepackage{bm}
\usepackage{hyperref}
\usepackage{stackrel,amssymb}
\usepackage{placeins}
\hypersetup{colorlinks=true, linkcolor=blue, citecolor=blue, urlcolor=magenta, pdftitle={On the search of displacive chiral phase transitions}, pdfauthor={F. Gomez-Ortiz}}
\usepackage{bbding}
\usepackage{xcolor} 
\usepackage[normalem]{ulem}

\begin{document}
\preprint{APS/123-QED}
\title{Multimodal Topological Textures Arising from Coupled Structural Orders in SrTiO$_3$}
\author{Fernando G\'omez-Ortiz}
\email[Corresponding author:]{fgomez@uliege.be}
\affiliation{Theoretical Materials Physics, Q-MAT, Université de Liège, B-4000 Sart-Tilman, Belgium} 
\author{Louis Bastogne}
\affiliation{Theoretical Materials Physics, Q-MAT, Université de Liège, B-4000 Sart-Tilman, Belgium}
\author{Philippe Ghosez}
\email[Corresponding author:]{Philippe.Ghosez@uliege.be}
\affiliation{Theoretical Materials Physics, Q-MAT, Université de Liège, B-4000 Sart-Tilman, Belgium} 

\date{\today}
\begin{abstract}
Magnetic spin topological textures recently found their electrical counterparts in polar topologies emerging from the condensation of inhomogeneous polar atomic distortions. Here, we further extend the concept to other non-polar atomic  degrees of freedom. Taking SrTiO$_3$ as a prototypical example, we investigate from second-principles atomistic simulations,  the equilibrium domain structures and topological textures associated with the natural antiferrodistortive rotations  of its oxygen octahedra.
Besides the common 90$^\circ$ antiferrodistortive domain walls (twin boundaries), we identify new metastable 180$^\circ$ domain walls oriented along the $\lbrace100\rbrace_\mathrm{pc}$ direction, when compressive epitaxial strain is applied. These domains exhibit complex antiferrodistortive Bloch- and Néel-like configurations with the later being the most favorable. We also stabilize antiferrodistortive vortex and antivortex structures which are accompanied by co-localized polarization vortices and a complex pattern of the local strain field, giving rise to a trimodal topological structures. Our results extends the concept of topological ordering to non-polar structural degrees of freedom and highlights the role of lattice-mediated couplings in stabilizing complex textures in perovskite oxides.
\end{abstract}
\maketitle
Topology has become a cornerstone of modern condensed matter physics, providing a powerful framework for understanding many physical systems~\cite{Mermin-79}. The concept of topological defects and their quantification through invariants aids to classify and predict the properties of materials in many different contexts, extending the usage of topology from abstract mathematical theory to tangible physical systems with profound implications for technology~\cite{Mermin-79}. 
Many examples can be encountered ranging from the role of dislocations in the deformation and melting of crystals~\cite{Kosterlitz-82,Anderson-17} to vortex-lines in the order parameter of superconductors and superfluids~\cite{Abrikosov-04,Anderson-17} or Chern numbers and robust edge states in topological insulators leading to new transport phenomena~\cite{Hasan-10}.  

Magnetic systems were also among the first solid-state platforms to host topologically nontrivial spin textures such as complex domain walls, vortices, and skyrmions~\cite{Runge-96,Shinjo-00,Robler-06,Nagaosa-13}. These objects, arising as continuous modulations of the magnetization vector field, are stabilized by competing interactions. They can exhibit particle-like behavior and show interesting functional applications~\cite{Stuart-08,Fert-13,Jibiki-20,Pinna-18}.
Since the beginning of this century, their polar counterparts have become a popular topic of research as well~\cite{Catalan-12,Junquera-23,LUKYANCHUK2025} and have been stabilized and characterized both theoretically and experimentally. 

Despite these specific advances, topological textures still remain marginally explored, and for instance, have not been explicitly considered in the context of other structural order parameters. In particular, the antiferrodistortive rotation patterns of oxygen octahedra, which constitute a primary structural order parameter in many perovskites can also be considered as a potential host for topologically nontrivial spatial modulations~\cite{Huang-14,Huang2016}. While oxygen octahedra rotations (OOR) are typically viewed as zone-boundary single-$q$ distortions, they in fact define continuous, unit-cell-resolved vector fields, in which multiple $q$-modulations can emerge much like in the case of inhomogeneous polar textures where multiple $q$-point modulations appear beyond a simple $\Gamma$ distortion. As such, they can, in principle, support topological features, including complex domain wall geometries~\cite{Eliseev-19} or vortices~\cite{Huang2016}.
\begin{figure}[b]
    \centering
      \includegraphics[width=\columnwidth]{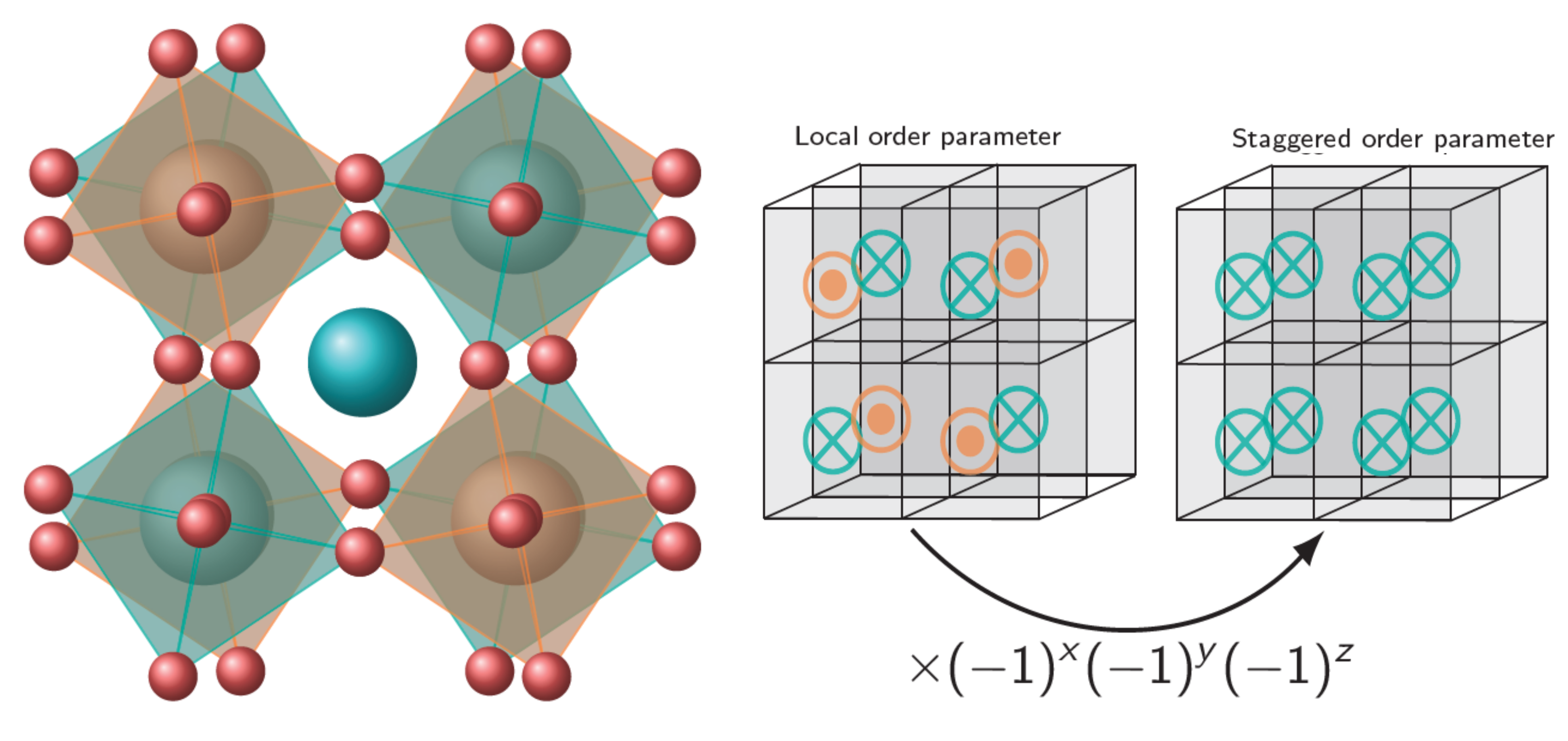}
      \caption{Oxygen octahedra rotations following an out-of-phase antiferrodistortive mode along an axis perpendicular to the plane of the paper. The pristine and staggered local rotations of the TiO$_6$ octahedra are shown.}
      \label{fig:Approts} 
\end{figure}
Similarly, strain fields in ferroelastic materials~\cite{Ekkard-12} can be described by a vector field indicating the local strain direction and are thus also subject to exhibit topological ordering~\cite{Nataf-20}.
\begin{figure*}[thbp]
     \centering
      \includegraphics[width=16cm]{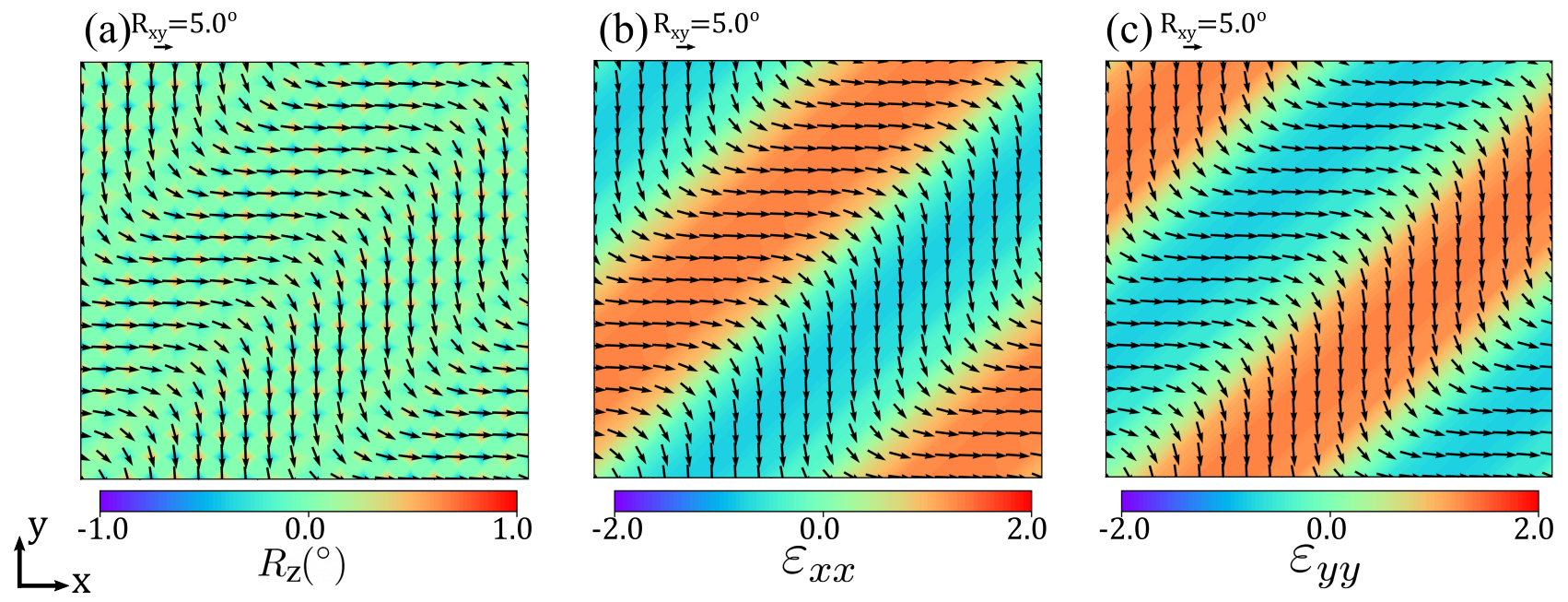}
      \caption{Relaxed 90$^\circ$ domain wall along the $\lbrace110\rbrace_{\rm pc}$ direction of the oxygen octahedral rotations in SrTiO$_3$. Arrows indicate the in-plane local rotations in degrees accounting with the phase shift explained on the main text where as colors indicate (a) the out of plane rotations (b) the local strains in \% along $x$ and (c) the local strains in \% along $y$.}
      \label{fig:figure1} 
\end{figure*}
Beyond their intrinsic interest, topological textures in OOR or other structural order parameters may induce the emergence of multimodal topological states, in which topological defects of different structural order parameters co-exist and are coupled together~\cite{Ekkard-20,Benedek-11,Haeni-04}. 
In perovskite oxides, OOR are often intimately coupled to polar distortions~\cite{He-24,Stengel-17,Bousquet-08} or to magnetic ordering~\cite{Benedek-11,Mulder-13}. 
As a result, a topological defect in the OOR-field, could induce or stabilize a correlated defect in the polarization or spin texture, giving rise to multimodal topological structures with multifunctional responses. 

In this work, considering SrTiO$_3$ as a prototypical example, we demonstrate that the OOR of bulk perovskites can exhibit emergent topological textures; including complex domain walls similar to those encountered in ferroelectrics~\cite{Wojdel-14} as well as vortices and antivortices, when subject to suitable mechanical boundary conditions. 
Furthermore, we show that these OOR-vortices are intrinsically coupled with vortex-like textures in both the polarization ($P$) and the local strain fields ($\varepsilon$), giving rise to multimodal (triple-) topological structures.
This view extends the landscape of topological phenomena to other structural order parameters and suggests new routes to control and exploit structural degrees of freedom in complex oxides.

To explore these ideas quantitatively, we performed large-scale atomistic simulations based on second-principles models as implemented on \textsc{Multibinit}~\cite{gonze2020abinit}. Further computational details can be found in the End Matter.

In order to visualize antiferrodistortive OOR, we assign arrows representing the rotation of the TiO$_6$ octahedra, with the direction of each arrow indicating the local rotation axis and its length the magnitude of the rotation (typically in degree) as schematized in Fig.~\ref{fig:Approts}. 
Since OOR are antiferrodistortive motions involving anti-phase rotations in consecutive unit cells, we define a {\it staggered} rotation field by multiplying by a phase correction factor of $(-1)^x(-1)^y(-1)^z$ the rotation amplitude of the octahedra in cell ($x,y,z$). This transformation allows a uniform antiferrodistortive rotation pattern to appear as a monodomain state for clearer visualization. Note that this phase convention is origin-dependent, so that the plotted rotation field is only defined up to a global sign. 

For the polarization field we followed the regular approach where local polarizations (typically in $\mu$C/cm$^2$) are computed within a linear approximation as the product of the Born effective charge tensor in the reference cubic phase times the atomic displacements from the reference structure positions~\cite{meyer2002ab}. 
Finally, all the strains are computed in \% taking as a reference the cubic lattice parameter predicted for SrTiO$_3$ by the model. An arrow is then assigned to each point, with its components corresponding to the strain along the $xx$, $yy$ and $zz$ directions.

Let us start by analyzing the equilibrium domain structures of bulk SrTiO$_3$ obtained after full structural relaxation within the second-principles framework. To this end, we constructed 90$^\circ$ and 180$^\circ$ OOR-domain walls along various crystallographic orientations, considering the $\lbrace110\rbrace_{\rm pc}$ and $\lbrace100\rbrace_{\rm pc}$ planes which constitute a representative set of domain orientations up to symmetry. An overview of the different domain configurations considered is provided in Fig.~\ref{fig:Appfig1}.
After full structural relaxation (atomic coordinates and cell parameters) only the 90$^\circ$ domain walls along $\lbrace110\rbrace_{\rm pc}$ direction were found to be stable in good agreement with the experimental evidence~\cite{Hellberg-19,Gray-16}.

In Fig.~\ref{fig:figure1} we show the relaxed rotation pattern obtained for such 90$^\circ$ domain walls. The two domains are characterized by distinct orientations of the octahedral tilt axis, rotated by 90$^\circ$ with respect to each other.
\begin{figure*}[thbp]
     \centering
      \includegraphics[width=16cm]{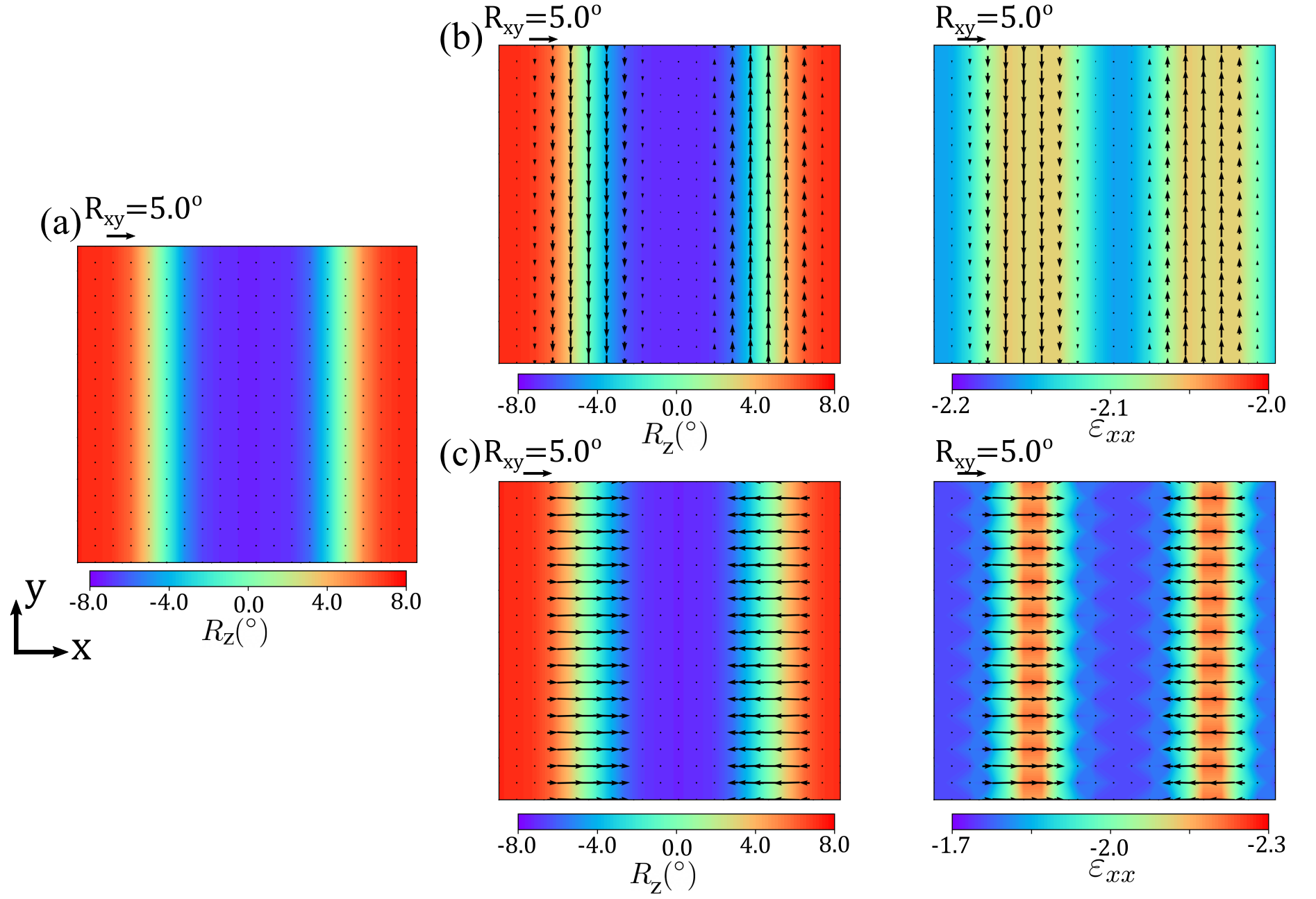}
      \caption{Relaxed (a) pure (b) Bloch and (c) Néel 180$^\circ$ domain wall along the $\lbrace100\rbrace_{\rm pc}$ direction of the oxygen octahedral rotations under -2.1\% epitaxially strained SrTiO$_3$. Arrows indicate the in-plane components of the oxygen octahedral rotations in degrees accounting with the phase shift explained on the main text whereas colors indicate either their out-of-plane component or the local strain relaxations along $x$ in \% as indicated on the labels.}
      \label{fig:figure2} 
\end{figure*}
The strong coupling between OOR and lattice strain gives rise to a ferroelastic domain wall~\cite{Salje_1991}, as evidenced by the local strain components $\varepsilon_{xx}$ and $\varepsilon_{yy}$ shown in panels (b) and (c), respectively. Regions with enhanced local strain along the $x$ (resp. $y$) direction correspond to domains where the OOR axis preferentially aligns along $x$ (resp. $y$), underscoring the intimate link between rotational order and ferroelastic distortion.
Interestingly, both head-to-head/tail-to-tail and head-to-tail configurations of the 90$^\circ$ domain wall were found to be equally stable within our calculations as shown in Fig.~\ref{fig:Appfig2}. This contrasts with the behavior of polarization domain walls, where bound charges and electrostatic considerations typically favor the head-to-tail configuration and suppress the formation of head-to-head or tail-to-tail domains. In the rotational case, however, both configurations coexist, reflecting the intrinsic symmetry nature of the OOR-order parameter. Such coexistence has been observed with first principles~\cite{Stengel-17} and the encountered domain wall thicknesses (1-2 u.c.) are consistent with the literature~\cite{Morozovska-12,Cao-90} further validating the accuracy of our second-principles approach for capturing the energetics and symmetry constraints of rotational domain configurations.

While all other domain configurations shown in Fig.~\ref{fig:Appfig1} were found to be unstable under full relaxation, we note that some of them can be stabilized under specific strain conditions in bulk SrTiO$_3$.
In particular, this is the case of 180$^\circ$ OOR domain walls oriented along the $\lbrace100\rbrace_{\rm pc}$ direction.
When subjected to a in-plane epitaxial macroscopic strain of $-2.1\%$ with respect to the cubic reference as explained in methods, the cell expands $1.5\%$ along the $z$-direction which becomes the preferred orientation for the OOR and these domains become stable.
The relaxed structure under this strain condition is shown in Fig.~\ref{fig:figure2} (a). As we can clearly see, without breaking the symmetry of the initial domain structure during the relaxation, sharp 180$^\circ$ domains are formed. The energy of the domain wall is of $182.8$ meV$/\square$, which is close to the typical values encountered for ferroelectric domains~\cite{meyer2002ab,Wojdel-14,Chege-25}.
Very interestingly, if the symmetry of the initial domain configuration is explicitly broken prior to relaxation, the system can stabilize more complex domain wall structures showing either Bloch [Fig.~\ref{fig:figure2}(b)] and Néel [Fig.~\ref{fig:figure2}(c)] type behaviors.
Remarkably, both Bloch and Néel configurations are stable and both lower the energy compared to the pristine sharp $180^\circ$ domain.
\begin{figure*}[thbp]
     \centering
      \includegraphics[width=\textwidth]{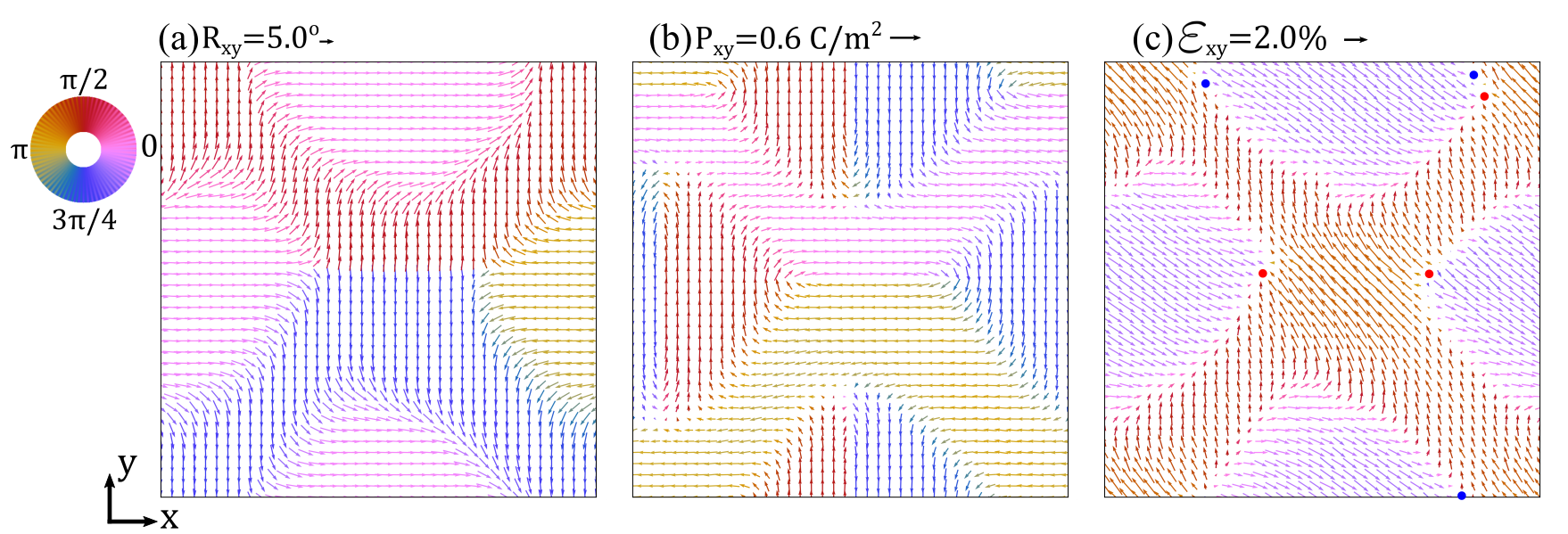}
      \caption{Fully relaxed vortex texture on a 40$\times$40$\times$2 supercell under no mechanical boundary conditions. (a) Oxygen octahedral rotation pattern, (b) polarization pattern and (c) local strain pattern. Arrows indicate the in-plane components of the corresponding vector field. The color wheel encodes their in-plane orientation, providing a complementary visualization to the arrows. The red and blue dots in (c) serve as visual guides to help identify the positions of some vortices and antivortices in the local strain field. }
      \label{fig:figure3} 
\end{figure*}
Among the two rotational textures, the Néel-type configuration is slightly lower in energy ($153.4$ meV$/\square$) than the Bloch-type wall ($178.5$ meV$/\square$). This energy difference can be rationalized by examining the local strain maps. At the domain wall, the local strain along the $z$-direction is reduced, while the strain along the $x$-direction is slightly relaxed, showing less negative values. In contrast, the strain along the domain wall $y$-direction remains homogeneous, with a constant value of $-2.1\%$.
Due to the strong coupling between strain and OOR, the Néel-type wall, where the dominant rotation axis lies along $x$, is better accommodated under the imposed strain conditions. Conversely, in the Bloch-type wall, the dominant rotation axis lies along the $y$-direction, which is subject to stronger compressive strain, resulting in a less favorable energetic configuration. 

Having established the stability and energetics of individual 90$^\circ$ and 180$^\circ$ OOR-domain walls under different strain conditions, we now turn our attention to more intricate structures that arise from the combination of these simple textures.
Let us start by discussing OOR-vortex/antivortex configurations constructed by the combination of 90$^\circ$ domains. 
Figure~\ref{fig:figure3} displays the relaxed structure of an OOR-antivortex on a $40\times40\times2$ supercell, formed by the appropriate arrangement of four distinct 90$^\circ$ OOR-domains as shown in Fig.~\ref{fig:Appfig3} (a).
As it is clearly shown in Fig~\ref{fig:figure3} (a), an extended OOR-antivortex is formed at the center of the supercell with convergent $x$-components and divergent $y$-components of the OOR.
As a consequence of the periodic boundary conditions, a OOR-vortex texture can also be envisioned at the $x$-lateral edge of the supercell that forms to compensate the overall vorticity of the structure. In Fig.~\ref{fig:Appfig3} (b) we show the pattern at the center of the supercell for visualization purposes after applying a translation.

Interestingly, the vortex-antivortex structure in the OOR-order parameter is accompanied by a co-localized vortex-like texture in the polarization field. As shown in Fig.~\ref{fig:figure3}(b), the polarization vector field exhibits extended $P$-vortices of comparable spatial extent to those observed in the OOR, suggesting a strong coupling between the two order parameters. 
Moreover, both OOR vortices and antivortices are consistently associated with $P$-vortices, while $P$-antivortices appear only in the interstitial regions between these OOR topological defects.
In addition, the local strain field also displays a non-trivial texture showing $\varepsilon$-vortices and antivortices as shown in Fig.~\ref{fig:figure3}(c) . However, in contrast to the extended nature of the OOR and $P$ textures, the singularities in the strain field are confined to isolated unit cells, indicating a different characteristic length scale.
This coexistence of OOR, $P$ and $\varepsilon$ vortex-like configurations reveal a complex, multimodal topological state involving three coupled structural order parameters. In such systems, the strain field couples linearly to the biquadratic invariants of both $P$ and OOR~\cite{Salje_1991,Zubko-07,Tagantsev-01}, while additional couplings between $P$ and gradients of OOR further stabilize the nontrivial textures by rotopolar couplings~\cite{Stengel-17}. These interactions highlight the intricate interplay between structural, polar, and elastic degrees of freedom.

Finally, inspired by the previous strategy and the approach followed in Ref.~\cite{Mauro-19,Gomez-25} to engineer ferroelectric skyrmions, we attempted to stabilize a OOR-nanodomain and a OOR-skyrmion-like texture in SrTiO$_3$ by combining 180$^\circ$ OOR-domain walls under appropriate strain conditions.
Surprisingly, despite the fact that the OOR-domain wall energy is comparable to that of ferroelectric systems where skyrmions have been successfully stabilized numerically, the constructed OOR-nanodomain was found to be unstable, collapsing upon relaxation into a monodomain state with uniform OOR.
Similarly, attempts to construct a full OOR-skyrmion texture by arranging multiple 180$^\circ$ OOR-Néel walls also led to collapse of the nanoregion. The Néel configuration was chosen as they have shown the lowest-energy arrangement for 180$^\circ$ OOR-domain boundaries. However, Bloch configurations were also tested yielding to similar results.
These results indicate that, under the conditions explored (strain, geometry and orientation of the nanocolumn), the OOR order parameter in SrTiO$_3$ does not support stable skyrmion-like configurations.

In summary, we have systematically investigated the equilibrium antiferrodistortive domain structures in bulk SrTiO$_3$ via second-principles atomistic simulations. We have focused on both, individual domain structures and topological textures constructed from the combination of the formers.
Full structural relaxations revealed that only the 90$^\circ$ OOR-domain walls along $\lbrace110\rbrace_{\rm pc}$ are stable under unstrained bulk conditions, consistent with experimental observations and previous first-principles predictions. However, under compressive epitaxial strain, we demonstrated the stabilization of 180$^\circ$ OOR-domain walls along the $\lbrace100\rbrace_{\rm pc}$ direction. Similarly to the ferroelectric case, these OOR-domain walls present a complex structure within the domain wall that is able to accommodate Bloch and Néel like configurations. The later were shown to be the most stable configuration due to the coupling between oxygen octahedra rotations and local strain relaxations.

Beyond isolated domains, we constructed more complex textures, such as OOR-vortex–antivortex pairs formed by the junction of four 90$^\circ$ OOR domains. Remarkably, the OOR-vortices are accompanied by co-localized $P$-vortices with a similar spatial distribution, as well as by $\varepsilon$-vortex–antivortex textures in the strain field, resulting in a multimodal topological texture involving OOR, $P$, and $\varepsilon$ degrees of freedom.
Finally, we attempted to stabilize OOR-nanodomains and skyrmion-like configurations under strain, but our simulations did not yield any metastable solutions, suggesting that either different boundary conditions, geometries or orientations of the domains may be required to realize such topological objects in bulk SrTiO$_3$.

Our results extend the concept of topological textures beyond ferroelectricity and ferromagnetism, demonstrating that antiferrodistortive, strain fields or other structural order parameters can also support rich, stable, and coupled topological configurations. Beyond their intrinsic interest, these novel topological textures can constitute a fertile playground to tune thermal conductivity properties~\cite{Nataf-20} as well as polar and magnetic ordering of materials~\cite{Stengel-17,Bousquet-08,Benedek-11}.
\acknowledgments
F.G.O. acknowledges financial support from MSCA-PF 101148906 funded by the European Union and the Fonds de la Recherche Scientifique (FNRS) through the grant FNRS-CR 1.B.227.25F and the Consortium des Équipements de Calcul Intensif (CÉCI), funded by the F.R.S.-FNRS under Grant No. 2.5020.11 and the Tier-1 Lucia supercomputer of the Walloon Region, infrastructure funded by the Walloon Region under the grant agreement No. 1910247. F.G.-O. and Ph. G. also acknowledge support by the European Union’s Horizon 2020 research and innovation program under Grant Agreement No. 964931 (TSAR).  Ph. G. also acknowledge support from the Fonds de la Recherche Scientifique (FNRS) through the PDR projects PROMOSPAN (Grant No. T.0107.20) and TOPOTEX (Grant No. T.0128.25).
%
\clearpage
\onecolumngrid
\section*{End Matter}
\twocolumngrid
\renewcommand{\thefigure}{A\arabic{figure}}
\setcounter{figure}{0} 
\emph{Computational details.-}
Second-principles (SP) atomistic models were constructed using the \textsc{Multibinit}~\cite{gonze2020abinit} software by fitting data from density functional theory (DFT) calculations produced with the \textsc{Abinit}~\cite{gonze2020abinit} software package. The generalized gradient approximation (GGA) with the PBESol exchange-correlation functional and a planewave-pseudopotential approach with optimized norm-conserving pseudopotentials from the PseudoDojo server~\cite{hamann2013optimized,van2018pseudodojo} were employed, considering as valence electrons $4s^{2}4p^65s^2$ for Sr, $3s^23p^63d^24s^2$ for Ti, and $2s^22p^4$ for O. 
%
The harmonic part of the models was derived from Density Functional Perturbation Theory (DFPT) as implemented in \textsc{Abinit}~\cite{gonze2020abinit}. Dynamical matrices for the relaxed cubic $Pm\bar{3}m$ structure were computed using a $8\times 8\times 8$ q-point mesh. Various properties such as dipole-dipole interaction, Born effective charge, strain-phonon coupling, and elastic constants were extracted from the DFPT framework.
The anharmonic part of the SP potential was fitted from DFT configurations. A cutoff radius of $\sqrt{3}/2$ times the cubic lattice cell parameter was used to generate the anharmonic symmetry-adapted terms (SAT) from third to eighth order, considering strain-phonon coupling and anharmonic elastic constants.

Structural relaxations were carried out using a combination of Hybrid Molecular Dynamics-Monte Carlo approach~\cite{duane1987hybrid,betancourt2017conceptual} and Broyden-Fletcher-Goldfarb-Shanno (BFGS) algorithm as implemented in {\sc{abinit}}~\cite{gonze2020abinit}.
\begin{figure}[tbhp]
     \centering
      \includegraphics[width=\columnwidth]{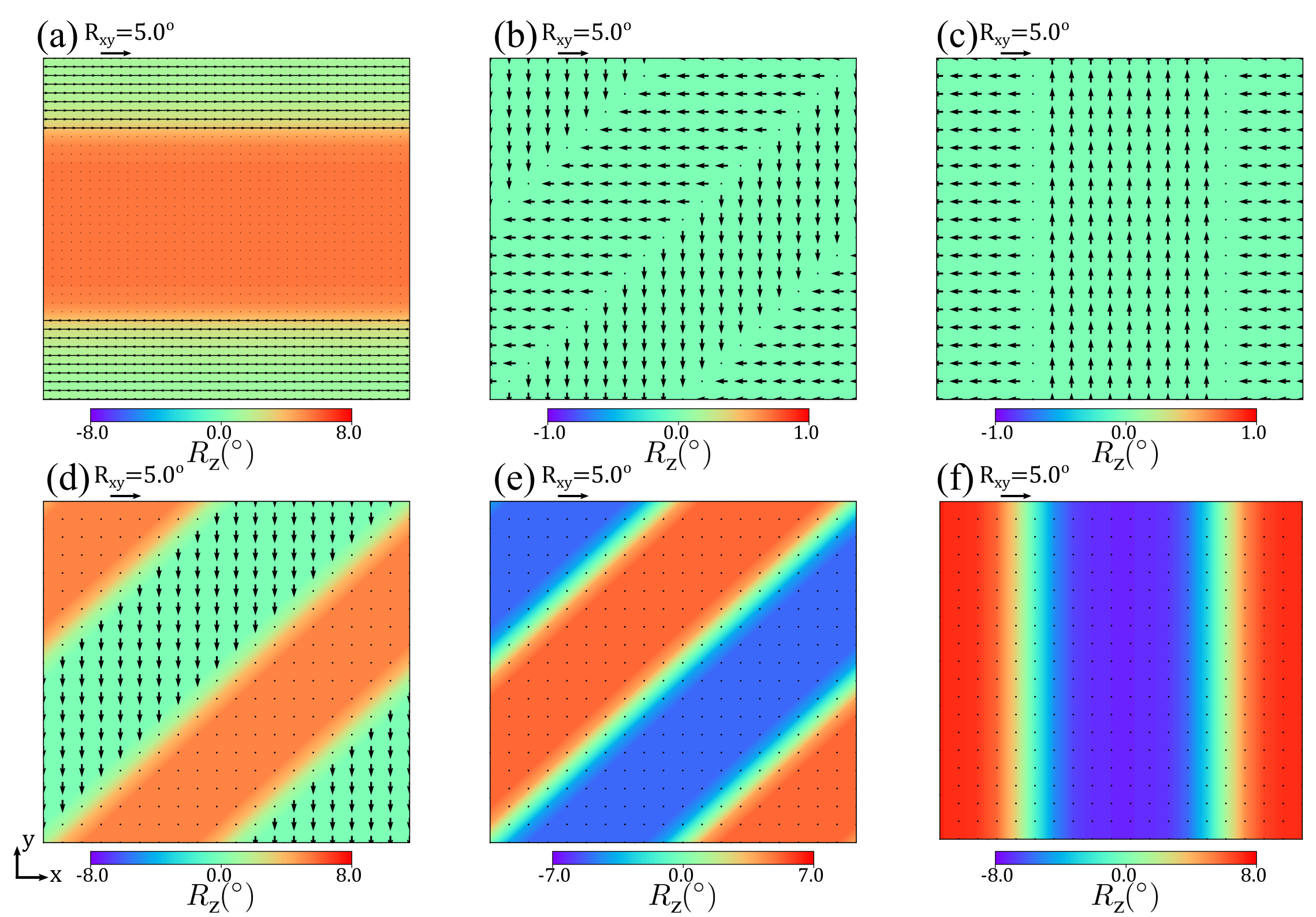}
      \caption{Different initial geometries for the domain walls studied on the work. Arrows indicate in-plane components of the oxygen octahedra rotations with the corresponding phase shift as explained in the main text whereas color bars indicate their out of plane component.}
      \label{fig:Appfig1} 
\end{figure}

\emph{Domain Structures.-}
In this section, we present the various domain structures investigated in this work. Figure~\ref{fig:Appfig1} displays the different initial domain geometries considered prior to structural relaxation. We tested a representative set of domains with diverse orientations and geometries to explore their stability as discussed on the main text.
Moreover, for the 90$^\circ$ domain wall we tested both, the head-to-head/tail-to-tail and the head-to-tail configurations.  Contrary to the case of polarization domains, these OOR-domain configurations are degenerate in energy.
\begin{figure}[thbp]
     \centering
      \includegraphics[width=8cm]{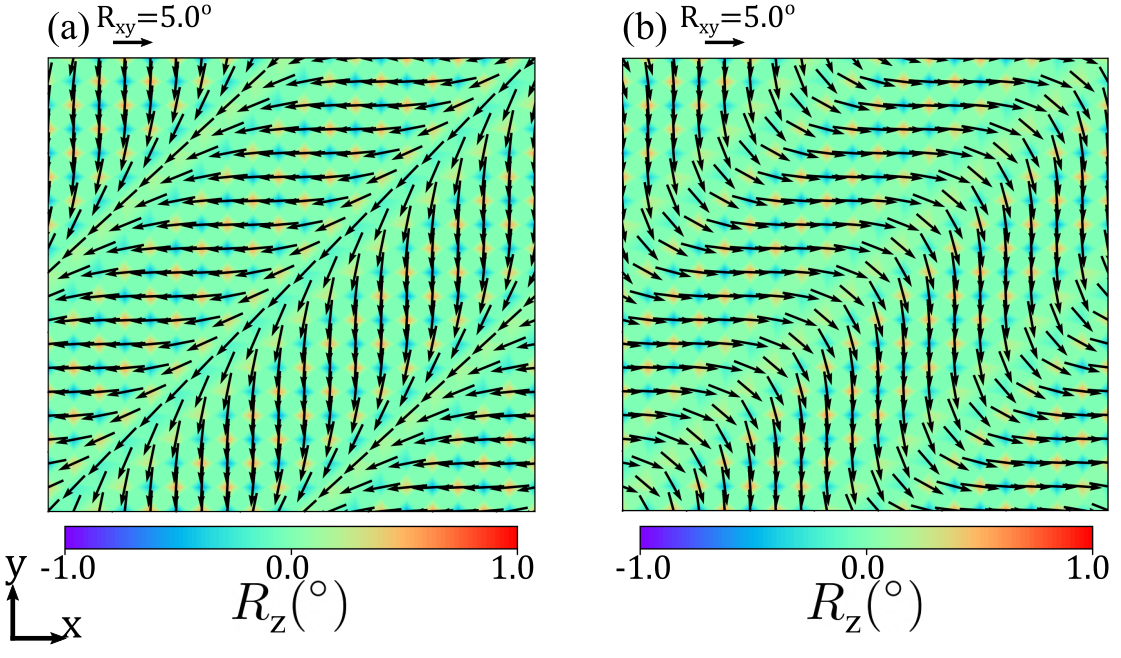}
      \caption{Different relaxed domain configurations for the 90$^\circ$ domain with $\lbrace110\rbrace_{\rm pc}$ orientation where oxygen octahedra rotations present either a head-to-head/ tail-to-tail character or a head-to-tail character.}
      \label{fig:Appfig2} 
\end{figure}

\begin{figure}[thbp]
     \centering
      \includegraphics[width=\columnwidth]{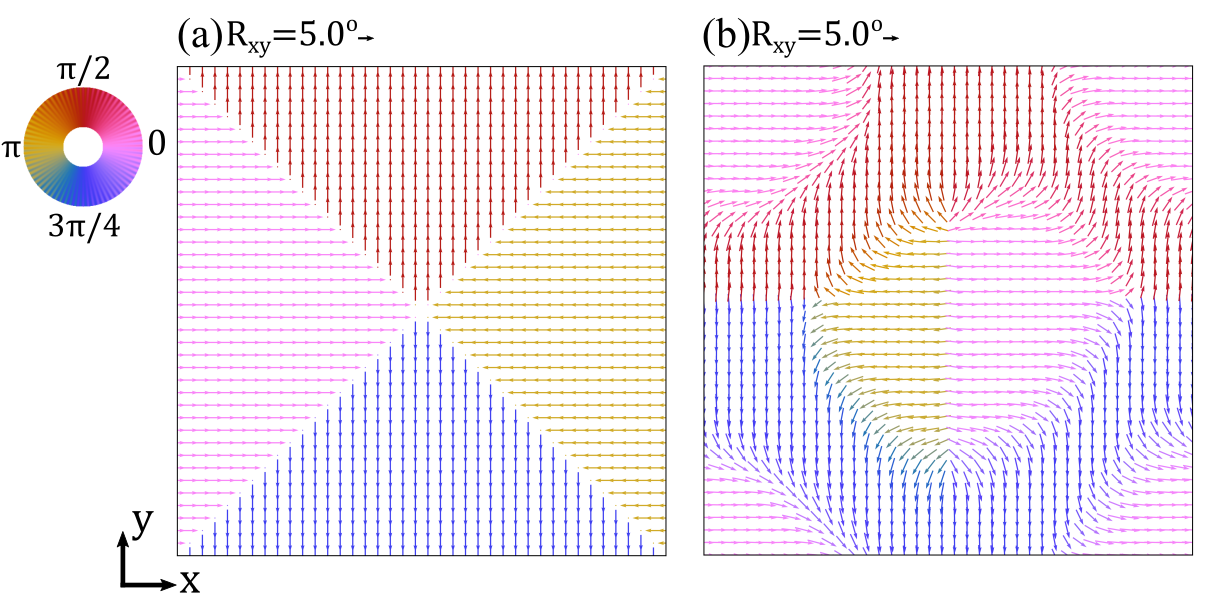}
      \caption{(a) Initial vortex texture of the oxygen octahedra rotations used for the relaxation of the structure presented in Fig.~\ref{fig:figure3} (a). (b) Vortex structure presented in Fig.~\ref{fig:figure3}(a) translated by 20 unit cells along the $x$-direction. Arrows and colors as in Fig.~\ref{fig:figure3} (a).}
      \label{fig:Appfig3} 
\end{figure}

Combining several simple domain structures we can build different textures such as vortices or antivortices. For instance, combining four 90$^\circ$ domain walls we can condense the structure represented on Fig.~\ref{fig:Appfig3} (a) that after relaxation lead to the vortex texture presented on Fig.~\ref{fig:figure3}(a).
Due to the periodic boundary conditions antivortices and vortices are balanced in the equilibrium structure, in our particular case, vortices are formed at the $x$-edge of the simulation cell as discussed on the main text. In Fig.~\ref{fig:Appfig3}(b) we show the same vortex structure presented on the main text translated by half a supercell vector to place the vortex at the center for visualization purposes.
\end{document}